\documentclass{JHEP3}

\newcommand{\tfrac}[2]{{\textstyle\frac{#1}{#2}}}

\title{The Higgs field and the ultraviolet behaviour
of the vortex operator in 2+1 dimensions}
\author{Arsen Khvedelidze\,$^{a,b}$
,\ Alex Kovner\,$^{a,c}$\
and David McMullan\,$^{a}$ \\ $^{a}$School of Mathematics and Statistics,
University of Plymouth, Plymouth PL4 8AA, UK\\$^{b}$Joint Institute for
Nuclear Research,  Dubna 141980, Russia\\$^{c}$Physics Department,
University of Connecticut, 2152 Hillside Road,
Storrs,\\\hspace*{.1cm} CT 06269-3046, USA\\
E-mail: \email{akhvedelidze@plymouth.ac.uk},
\email{akovner@plymouth.ac.uk},
\\ \hspace*{1.15cm}  \email{dmcmullan@plymouth.ac.uk}}

\abstract{We calculate the change in the ultraviolet behaviour of
the vortex operator due to the presence of dynamical Higgs field
in both 2+1 dimensional QED and the 2+1 dimensional Georgi-Glashow
model. We find that in the QED case the presence of the Higgs
field leads at the one loop level to power like correction to the
propagator of the vortex operator. On the other hand, in the
Georgi-Glashow model, the adjoint Higgs at one loop has no affect
on the vortex propagator. Thus, as long as the mass of the Higgs
field is much larger than the gauge coupling constant, the
ultraviolet behaviour of the vortex operator in the Georgi-Glashow
model is independent of the Higgs mass.}

\begin{document}

\section{Introduction}

During the last several years the vortex mechanism of confinement
has seen a steady increase in popularity. Originally introduced by
't\,Hooft in 1978 \cite{thooft} it has been for a long time
eclipsed by the rival monopole, or dual superconductor mechanism
\cite{mon}. The dual superconductor mechanism has attracted a lot
of attention during the 80's and most of the 90's. Nevertheless, a
steady background of work on the understanding of magnetic
vortices both on the lattice \cite{tomboulis} and in the continuum
\cite{kr} has been maintained. More recently the interest in the
vortex mechanism has been revived in the lattice community by the
series of works by Greensite and collaborators \cite{greensite}.
The notion that magnetic vortices are important for confinement
seems to have earned credibility even among the long time
proponents of the monopole condensation
mechanism~\cite{polikarpov}.

Ideally, of course, rather than just performing numerical studies,
one would like to have a simple theoretical description of the
relevant confining dynamics directly in terms of the operators
that create and annihilate magnetic vortices. While in 3+1
dimensions this goal has not been achieved, in 2+1 dimensional
confining theories such a description exists \cite{review}. The
effective low energy description in terms of the vortex operators
is well established in the weakly coupled confining theories in
2+1 dimensions. It provides a simple and intuitive picture of
confinement and also deconfining phase transition at finite
temperature \cite{deconf}.

It has been argued that the main properties of the vortex
operators and the confinement mechanism carry over to the strongly
interacting theories, like pure gluodynamics \cite{kr1}. The
reason is that in 2+1 dimensions the putative Higgs phase with a
perturbatively massless photon is not separated by a phase
transition from a confining phase. One therefore can imagine
taking the pure gluodynamics limit by starting with the weakly
coupled Georgi-Glashow model, where the confining physics is well
understood, and then continuously changing the coefficient of the
Higgs mass term so that it becomes positive and then very large.
In this transition between the Higgs regime and the confinement
regime one should not encounter any non-analyticity. It is still
far from straightforward to study the vortex operators in such a
theory, since their infrared physics is genuinely nonperturbative.
On the other hand, the ultraviolet physics in these theories is
perturbative. It is thus interesting to see what can be said about
the ultraviolet properties of the vortex operator, such as its
scaling behaviour at short distances, and in particular what are
the effects of decoupling the Higgs field on it.

This is the question we ask in this paper: \emph{what is the
influence of dynamical matter fields on the ultraviolet behaviour
 of the vortex operator
$V(x)$?} To answer this we study two 2+1 dimensional gauge
theories: noncompact quantum electrodynamics (QED) with a charged
scalar field, and the $SU(2)$ gauge theory with the Higgs field in
the adjoint representation. In the former theory confinement is
logarithmic rather than linear. Nevertheless the vortex operator
plays a crucial role in the low energy dynamics~\cite{kr}. The
latter theory in its weakly coupled incarnation is the simplest
known model which exhibits linear confinement between fundamental
charges~\cite{polyakov}. In this weakly coupled regime, where
classically the Higgs field has a nonvanishing expectation value,
this model is usually known as the Georgi-Glashow model. We are
interested in the opposite regime, namely when the VEV of the
Higgs vanishes and their masses become large. Nevertheless, since
the two regimes are analytically connected in 2+1 dimensions, we
will continue to call this theory the Georgi-Glashow model.

The regime of interest to us is when the mass of the Higgs field is much
larger than the gauge coupling constant ($M\gg g^2$), and the distance
between the vortex and the antivortex in the correlation function is
comparable to $M^{-1}$. We will see that the effect of the Higgs field is
quite different in the two models. In QED the dynamical Higgs field
affects the ultraviolet behaviour of the vortex operator at the one loop
level and leads to a power like factor in the vortex correlation function:
\begin{equation}
\left<V(x)V^*(y)\right>=\left({1\over|x-y|^2\Lambda^2}\right)^{1\over 8}
\left<V(x)V^*(y)\right>_0\,,
\end{equation}
where
\begin{equation}
\left<V(x)V^*(y)\right>_0=\exp\left(-{\pi\over
g^2}[\Lambda-{1\over|x-y|}]\right)
\end{equation}
is the vortex correlator in the theory without Higgs.

In the Georgi-Glashow model, on the other hand, it turns out that
the adjoint Higgs has no effect on the vortex correlator in this
regime. In this sense the pure gluodynamics regime in the
Georgi-Glashow model is achieved very efficiently, since the
moment the Higgs mass is large, the vortex correlation function
has its pure gluodynamics behaviour at all distance scales below
$g^{-2}$, including the true ultraviolet regime.

We start our discussion with the Abelian theory.

\section{The noncompact $U(1)$ model}

Consider the $U(1)$ gauge theory with a complex scalar matter
field defined by the (Euclidean) Lagrangian
\begin{equation}
L={1\over
4}F_{\mu\nu}^2+|(\partial_\mu+igA_\mu)\phi|^2+M^2\phi^*\phi\,.
\end{equation}
The construction of the vortex operator and its role in the low
energy dynamics of this theory  has been extensively reviewed
recently. The vortex operator is defined as
\begin{equation} \label{v}
V(x)=\exp\left\{{2\pi i\over g}\int_{C(x)}
E_i(y)\epsilon_{ij}dy_j\right\}\,,
\end{equation}
where the contour $C(x)$ starts at the point $x$ and goes to
infinity. The important properties of $V$ are:

i) $V(x)$ does not depend on the curve $C$;

ii) $V(x)$ is a local, gauge invariant scalar field.

\noindent These may not be immediately obvious  from the
definition eq.(\ref{v}), but nevertheless they have been
rigorously established (see \cite{review} for review).

The expectation value of $V$ in the Euclidean path  integral
formalism is given by the following expression
\begin{equation}\label{expv}
\left<V(x)\right>=Z^{-1}\!\!\int\! [dA_\mu] [d\phi]
\exp{-}\int\!\! d^3x \left({1\over
4}(F_{\mu\nu}-s_{\mu\nu})^2+|(\partial_\mu+igA_\mu)\phi|^2+M^2\phi^*\phi\right)\!,
\end{equation}
where  $Z$  is the normalization factor in the partition function
of the theory and the  $c$-number source function $s_{\mu\nu}$ is
given by
\begin{equation}
s_{\mu\nu}={2\pi\over
g}\epsilon_{\mu\nu\lambda}\tau_\lambda(C)\delta^2(x\in C)\,,
\end{equation}
with $\tau_\lambda(C)$  the unit vector tangent to the contour $C$.
 Due to the explicit presence of $1/g$ in the source,
 one can calculate the path integral eq.(\ref{expv}) in the
steepest descent approximation. To leading order the presence of
the Higgs field is irrelevant. The solution of the classical
equations of motion in the presence of the source $s_{\mu\nu}$ is
the Dirac monopole configuration with the Dirac string along $C$.
The action of this solution is
\begin{equation}
S_{{\rm classical}}={\pi\over 2g^2}\,\Lambda
\end{equation}
and  the corresponding vortex VEV is
\begin{equation}
\left<V\right>_0=e^{-{\pi\over 2 g^2}\,\Lambda}\,,
\end{equation}
where $\Lambda$ is the ultraviolet cutoff. The same calculation
for the vortex-antivortex correlation function gives
\begin{equation}
\left<V(x)V^*(y)\right>_0=e^{-{\pi\over
g^2}[\Lambda-{1\over|x-y|}]}\,.
\end{equation}

The dynamical Higgs field enters the calculation of the VEV at
one-loop resulting in
\begin{equation}\label{oneloop}
\left<V\right>=e^{-{\pi\over 2g^2}\,\Lambda}\, {\rm Det}^{-1}
\left({-|D|^2+M^2\over -\partial^2+M^2}\right)\,,
\end{equation}
where $D_\mu$ is the covariant derivative in the classical field
of a Dirac monopole. The rest of this section is devoted to the
calculation of the determinant in eq.(\ref{oneloop}).

The contribution due to the determinant is infrared finite, but
ultraviolet divergent.  Since the ultraviolet cutoff $\Lambda$ and
the Higgs mass are  the only scales in the calculation, the result
must be
\begin{equation}\label{q}
\ln{\rm Det} \left({-|D|^2+M^2\over -\partial^2+M^2}\right)={\rm
tr}\ln \left({-|D|^2+M^2\over -\partial^2+M^2}\right)=\alpha\ln
\left({\Lambda^2\over M^2}\right)\,,
\end{equation}
with $\alpha$  a pure number and we neglect corrections with
positive powers in $M/\Lambda$. Our aim is to calculate this
number $\alpha$. For convenience we will  consider the derivative
of eq.(\ref{q}) with respect to $M^2$:
\begin{equation} \label{q1}
{\rm tr}\left[{1\over -|D|^2+M^2}-{1\over
-\partial^2+M^2}\right]=-\alpha M^{-2}\,.
\end{equation}

This  calculation is equivalent to solving  the quantum mechanical
problem of a scalar particle in the field of the Dirac monopole of
unit magnetic charge. Therefore, consider a quantum mechanical
Hamiltonian
\begin{equation}
H=-D^*D+{\mathcal{V}}(r)
\end{equation}
and the associated time-independent Schr\"odinger equation
\begin{equation}
H\Psi=\epsilon\Psi\,.
\end{equation}

For any rotationally invariant potential ${\mathcal{V}}(r)$ the
angular part of the problem is solved by separation of variables:
\begin{equation}
\Psi=f_l(r)Y^q_{l,m}(\theta,\phi)\,,
\end{equation}
where $Y^q$ are the so-called monopole harmonics  \cite{monoTamm},
\cite{monoWuYang} corresponding to a magnetic monopole of magnetic charge
$4\pi q/g$ and are analogous to the ordinary spherical functions. In our
case $q=1/2$ but we shall keep $q$ general for the moment. The angular
momentum quantum number $l$ takes the values $l=q,\ q+1,\ \dots $ and the
magnetic quantum number $m=-l,-l+1, \dots,l$.

Although the final result eq.(\ref{q1}) is infrared finite, each
one of the two terms has a part proportional to the volume. It is
thus convenient to introduce an infrared regulator. We choose to
do this by placing our quantum mechanical system in the potential
of a spherically symmetric harmonic oscillator
\begin{equation} \label{harm}
{\mathcal{V}}(r)={1\over 2}\ \omega r^2\,.
\end{equation}
The thermodynamic limit is recovered as $\omega\rightarrow 0$.

The  radial wave function $f_l(r)$ satisfies the equation
\begin{equation} \label{rad}
\left[-\left({\partial^2\over \partial r^2}+{2\over
r}{\partial\over\partial r}\right)+{l(l+1)-q^2\over r^2}+{1\over 2}\
\omega r^2\right]f_l(r)=\epsilon f_l(r)\,.
\end{equation}
The spectrum of eq.(\ref{rad}) can be found in a similar way to
the well known case of the ordinary spherically symmetric
oscillator problem (see e.g. \cite{davydov}). Requiring  that the
radial function $ f_l(r) $ vanishes at infinity, we have:
\begin{equation}
\epsilon^{(\pm)}_{n,l}=\omega(2n+2\rho_l^{(\pm)}+\tfrac12)\,,
\end{equation}
with integers $n=0,1,2,\dots$ and parameters
\begin{equation}\label{eq:spb}
\rho_l^{(\pm)}={1\over 4}\pm{1\over 2}\sqrt{(l+\tfrac12)^2-q^2}\,.
\end{equation}
The parameters $\rho_l^{(\pm)}$ control the behaviour  of the radial wave
function $f_l$ at the origin:
\begin{equation}\label{eq:rfbehav}
    f_l (r)\ \propto \  r^{2\rho_l^{(\pm)}-1}\ \mathrm{as}\ r\to0\,.
\end{equation}
In order to ensure that the  Hamiltonian is self-adjoint, we
demand the boundary condition that $ f_l (r)$ is finite at the
origin. This requirement is satisfied only with the $\rho_l^{(+)}$
branch of the solutions (\ref{eq:spb}). Therefore we  take as the
spectrum for our problem $\epsilon_{n,l}=\epsilon^{(+)}_{n,l}$ and
consider the expression for the powers of the resolvent
\begin{equation} \label{eq:sum}
{\rm tr}\left({1\over
-|D|^2+M^2}\right)^s=\sum_{n=0}^{\infty}\sum_{l=q}^{\infty}{2l+1\over
[\epsilon_{n,l}+M^2]^s}\,,
\end{equation}
where an appropriate power $s$ has been introduced so as to ensure
the absolute convergence of the infinite sums at intermediate
stages in the calculation. Here the factor $2l+1$ in the numerator
is the degeneracy due to the magnetic quantum number $m$. We can
rewrite eq.(\ref{eq:sum}) as
\begin{equation} \label{sum}
\sum_{n=0}^{\infty}\sum_{l=q}^{\infty}{2l+1\over
[\epsilon_{n,l}+M^2]^s}={2\over
(2\omega)^s}\sum_{n=0}^{\infty}\sum_{l=0}^{\infty}{(l+{1\over
2})(1+x)\over \left[n+{1\over 2}+{1\over 2}(l+{1\over
2})\sqrt{1+2x}+{M^2\over2\omega}\right]^s}\,,
\end{equation}
with
\begin{equation}
x={q\over l+{1\over 2}}\,.
\end{equation}

For our purposes it is convenient to expand this expression in
powers of $x$. We only need to keep terms up to and including
order $x^3$. The reason is that each power of $x$ makes the
summation over $l$ and $n$ more convergent. Thus an expansion in
powers of $x$ under the summation sign is related to the expansion
of the result of the summation in powers of $\omega/ M^2$. In
fact, as we shall see below, each additional power of $x$ leads to
at least one additional power of $\omega/M^2$. The term of order
$x^0$ results in the sum of order $\omega^{-3}$ (this term
diverges in the limit $\omega\rightarrow 0$ and has to cancel
against the second term in eq.(\ref{q1})). Thus all terms starting
with $x^4$ vanish in the thermodynamic limit $\omega\rightarrow
0$. The only relevant terms in eq.(\ref{sum}) are therefore
\begin{eqnarray}\label{sum1}
{2\over
(2\omega)^s}\sum_{n=0}^{\infty}\sum_{l=0}^{\infty}&&{(l+{1\over
2})\over \left[n+{3\over 4}+{l\over 2}+{M^2\over
2\omega}\right]^s}\left[1 +q\left(-{s\over 2[n+{3\over 4}+{l\over
2}+{M^2\over 2\omega}]}+{1\over l+{1\over 2}}\right)\nonumber\right.\\
&&\left. +q^2\left(-{s\over 4[n+{3\over 4}+{l\over 2}+{M^2\over
2\omega}](l+{1\over 2})}+{s(s+1)\over 8[n+{3\over 4}+{l\over
2}+{M^2\over 2\omega}]^2}\right)\right.\\
&&\qquad \left.+ q^3\left(-{{s(s+1)(s+2)\over 48 [n+{3\over 4}+{l\over
2}+{M^2\over 2\omega}]^3}}\right)\right]\,.\nonumber
\end{eqnarray}
Note that, as expected,  the first term in this expression exactly cancels
the second term  in eq.(\ref{q1}). To calculate the rest of the terms we
use the following two basic integrals
\begin{eqnarray}
\sum_{n=0}^{\infty}\sum_{l=0}^{\infty}{(l+{1\over 2})\over
\left[n+{3\over 4}+{l\over 2}+{M^2\over
2\omega}\right]^p}&=&{1\over 16}{1\over \Gamma(p)}\int_0^\infty
dt\, t^{p-1}e^{-{M^2\over
2\omega}t}{1\over \sinh^3({t\over 4})}\,,\\
\sum_{n=0}^{\infty}\sum_{l=0}^{\infty}{1\over \left[n+{3\over
4}+{l\over 2}+{M^2\over 2\omega}\right]^p}&=&{1\over 8}{1\over
\Gamma(p)}\int_0^\infty dt \,t^{p-1}e^{-{M^2\over
2\omega}t}{1\over \cosh({t\over 4})\sinh^2({t\over
4})}\,.\label{formulae}
\end{eqnarray}
These expressions follows from the well-known infinite integral
representation for the generalized Riemann zeta function (see e.g.
\cite{WhittakerWatson}).

Expansion in powers of $\omega/M^2$ is simply achieved by
expanding the factors that multiply $\exp\{-{M^2\over 2\omega}t\}$
in the integrands in eqs.(\ref{formulae}) in powers of $t$.
Clearly increasing the power $p$ by one  or adding a factor of
$l+{1\over 2}$ in the denominator in these expressions, leads to
an extra power of $\omega/M^2$.  From eq.(\ref{sum1}), each
additional power of $q$ (and hence $x$) in the expansion is indeed
accompanied by such an increase and thus results in a higher power
of $\omega/M^2$ as claimed above. Some simple algebra shows that
the terms $O(\omega^{-2})$ and $O(\omega^{-1})$ in eq.(\ref{sum1})
vanish, and the final result to order $O(\omega^0)$ is
\begin{equation}
-\frac{1}{6}q(q^2+ {1\over 2})M^{-2s} \,.
\end{equation}
Now taking $s=1$,  $q =1/2 $ and referring back to eq.(\ref{q1}) we obtain
\begin{equation}
\alpha={1\over 16}\,.
\end{equation}

Putting these results together, we find that to one-loop order the
vacuum expectation value of the vortex operator is
\begin{equation}\label{vev1}
\left<V\right>=e^{-{\pi\over 2
g^2}\,\Lambda}\left({M^2\over\Lambda^2}\right)^{1\over 16}\,.
\end{equation}

Although we have not calculated directly the correlation function of the
vortex operator, eq.(\ref{vev1}) in conjunction with the mechanics of the
calculation allow us to understand its main features. The calculation of
the correlation function would lead to the quantum mechanical problem of a
particle in the background of the monopole-antimonopole pair separated by
a distance
 $r=|x-y|$. Clearly, if the
separation of the pair is larger than the inverse mass of the
Higgs  field, the result of the calculation will simply be the
square of
 eq.(\ref{vev1}). On the other hand, if the distance is much smaller
 that $M^{-1}$, the Higgs mass will not enter the final result.
 Instead the mass $M$ will be substituted by $r^{-1}$.
 Thus the correlation function has the form
\begin{equation}
\left<V(x)V^*(y)\right>=e^{-{\pi\over
g^2}[\Lambda-{1\over|x-y|}]}\left({M^2\over\Lambda^2}\right)^{1\over
8}f(M^2r^2)\,,
\end{equation}
where
\begin{eqnarray}
&&  \lim_{z\to\infty}f(z)=1\,,\\
&&  \lim_{z\to0}f(z) \,\, \propto \,\, z^{-1/8}\,.
\end{eqnarray}
We thus find that the presence of the dynamical Higgs field leads to an
additional factor $|x-y|^{-1/4}$ in the vortex-antivortex correlator in
the ultraviolet region.

\section{The $SU(2)$ Georgi-Glashow model}

Next we consider the $SU(2)$ gauge theory with an adjoint Higgs
field
\begin{equation}
L={1\over 4}(F^a_{\mu\nu})^2+(D^{ab}_\mu\phi^{b})^2+M^2\phi^a\phi^a\,.
\label{lgg}
\end{equation}
The vortex operator in this theory is defined as
\begin{equation}
V(x)=\exp\left\{{2\pi i\over g}\int_{C(x)}
E^a_i(y)\hat\phi^a(y)\epsilon_{ij}dy_j\right\} \label{vgg}
\end{equation}
with the unit vector
\begin{equation}
\hat\phi^a={\phi^a\over|\phi|}\,.
\end{equation}
As discussed in detail in \cite{review}, the theory is invariant
under the $Z_2$ magnetic symmetry, $V(x)\rightarrow -V(x)$. The
nonvanishing expectation value of $V(x)$ breaks this symmetry
spontaneously and the magnetic symmetry breaking is tantamount to
linear confinement.

Although the calculation of $\left<V\right>$ has been discussed in
some detail in the framework of the effective low energy theory
\cite{review}, we are not aware of its direct calculation in the
microscopic theory defined by the Lagrangian eq.(\ref{lgg}). We
therefore start our discussion by setting up this calculation. In
the path integral formalism
\begin{equation}\label{expvgg}
\left<V(x)\right>= Z^{-1} \int [dA^a_\mu][d\phi^a] \exp-\int d^3x \left(
{1\over4}(F^a_{\mu\nu}-\hat\phi^a
s_{\mu\nu})^2+(D\phi)^2+M^2\phi^2\right)\,.
\end{equation}

As in the case of QED, in the leading order the Higgs field is
irrelevant and we must minimize the pure Yang-Mills action in the
presence of the source $\hat\phi^as_{\mu\nu}$. To do this, we
first choose $\hat\phi^a=\delta^{a3}$. In the context of the
present calculation this can be considered as a gauge fixing.
Having found the classical solution for this source, $b^a_\mu$,
the solution for a general $\hat\phi$ is found by gauge
transforming it:
\begin{equation}
A^a_{\mu {\rm cl}}=U^\dagger b_\mu U+{i\over g}U^\dagger\partial_\mu U
\label{classg}\,,
\end{equation}
where the matrix $U$ is determined by the Higgs field via
\begin{equation}\label{eq:unh}
\sigma^a\hat\phi^a=U^\dagger\sigma^3 U
\end{equation}
with $\sigma^a$ the  Pauli matrices.

The problem of minimizing the action with the source is similar to
that in the Abelian theory. There it is known that the solution is
a pointlike Dirac monopole. In the present case this is not
entirely obvious. After all, we know that the $SU(2)$ theory has
finite action solutions for  monopoles with magnetic charge double
that of the elementary Dirac monopoles. If this also were true for
the case at hand, such a finite action solution would be preferred
over the pointlike Dirac monopole whose action is linearly
divergent in the UV. To put it another way, we know from the
effective theory calculations in the Higgs regime \cite{review},
that the leading behaviour of the VEV in eq.(\ref{expvgg}) is
$\exp\left(-2\pi{\lambda\over g^2}\right)$. What the effective
theory can not tell us is whether the scale $\lambda$ is the UV
cutoff of the effective theory (the mass of the $W$-boson $M_W$),
or the genuine UV cutoff $\Lambda$. The only way to settle this
question is to solve the classical equations of the underlying
$SU(2)$ gauge theory.

We will show now that the classical solution which minimizes the
action is the pointlike Dirac monopole, and that it has the action
which diverges linearly in the infinite cutoff limit.

It is useful to perform the calculation using the spherical
coordinates
\begin{equation}
x_1=r\cos\phi\sin\theta,\ \ \ x_2=r\sin\phi\sin\theta,\ \ \
x_3=r\cos\theta\,.
\end{equation}
It is known \cite{monopolesolutions} that for monopoles with even
magnetic charge classical solutions do not always have spherical
symmetry, but are rather axially symmetric. We thus take for our
solution the general (up to a gauge rotation around the third
axis) axially symmetric ansatz
\begin{equation} \label{ansatz}
g\ b^a_\mu\ \sigma^a = A(\theta,r)\partial_\mu \phi\
\sigma^1_{(n\phi)} + B(\theta,r)
\partial_\mu \theta \ \sigma^2_{(n\phi)}+ C(\theta)\ \partial_\mu\phi\
\sigma^3\,,
\end{equation}
where $A, B$ and $C$ are scalar functions, $n$ is an integer and
\begin{equation}
\sigma^1_{(n\phi)}=\cos(n\phi)\ \sigma^1+\sin(n\phi)\ \sigma^2\,,\
\ \ \sigma^2_{(n\phi)}=\cos(n\phi)\ \sigma^2-\sin(n\phi)\
\sigma^1\,.
\end{equation}
The dual field  strength $\tilde F^a_{\mu}= \frac{1}{2}
\epsilon_{\mu\nu\lambda}F^a_{\nu\lambda}$ for this ansatz is
\begin{eqnarray}\label{field}
 g \tilde F^a_{\mu}\ \sigma^a &=& \Big(\pi \delta_{\mu 3}[\Theta(\,x_3)C(0)+
\Theta(-x_3)C(\pi)] \delta(\,x_1)\delta(\,x_2)+ [C'-
AB]\epsilon_{\mu\nu\lambda}\partial_\lambda\phi
\partial_\nu\theta \Big)\,\sigma^3 \nonumber\\
 &&\qquad+
 \Big(\epsilon_{\mu\nu\lambda}\partial_\nu r\partial_\lambda\phi{\partial A\over\partial r}
 +\epsilon_{\mu\nu\lambda}
\partial_\nu\theta
\partial_\lambda\phi\,[A'+(n+C)B]\Big)\,\sigma^1_{(n\phi)}\\&&\qquad\qquad
+\epsilon_{\mu\nu\lambda}\partial_\nu r \partial_\lambda \theta\
{\partial B\over\partial r}\,\sigma^2_{(n\phi)}\,.\nonumber
\end{eqnarray}
Here $\Theta(x)$ is a step function and the prime over a function denotes
its derivative with respect to $\theta$. Clearly, in order to cancel the
string contribution in eq.(\ref{expvgg}) the function $C$ must satisfy
\begin{equation}\label{bound1}
C(\theta=0)=0\,,\qquad  C(\theta=\pi)=1\,.
\end{equation}
For the sake of generality we will for now take the string
contribution $s_{\mu\nu}$ in eq.(\ref{expvgg}) to have the
strength $\nu$, and thus take
\begin{equation}
C(\theta=\pi)=\nu\,.
\end{equation}
Eventually we are interested in $\nu=1$. In eq.(\ref{field}) we
have assumed that
\begin{equation}\label{bound2}
A(r,\theta=0)=0\,, \qquad A(r,\theta=\pi)=0\,,
\end{equation}
as otherwise there would be an extra string like singularity along
the $x_3$ axis in a $\sigma^1_{(n\phi)}$ component of the field
strength.

Note, that in order to have the action of this configuration ultraviolet
finite, one needs
\begin{equation}
\left[C'(\theta)-A(r, \theta)B(r, \theta)\right]\bigl|_{r=0}=0\,.
\end{equation}
However,  this condition  can be satisfied only for an even values of
$\nu$. Therefore, instead of imposing this condition we calculate the
action for the ansatz eq.(\ref{ansatz}) and minimize it with respect to
variational functions $A, B$ and $C$ as well as the integer parameter $n$.
The leading piece in the action is the UV divergent integral
\begin{equation}
S\sim\Lambda\, \int_{0}^{\pi}
\frac{d\theta}{\sin\theta}\{[C'-AB]^2+[B(C+n)+A']^2\}\,,
\end{equation}
where the functions $A$ and $B$ are evaluated at $r=0$. It turns
out that the minimization equations with respect to $A$ and $B$
can be solved exactly at given $C$ (see Appendix). The solution is
\begin{eqnarray}
&&B={AC'-A'(C+n)\over A^2+(C+n)^2}\,,\\
&&A^2+(C+n)^2=(1-a\cos\theta)^2b^2\,.
\end{eqnarray}
Imposing the boundary conditions at $\theta=0$ and $\theta=\pi$,
 we find that
\begin{eqnarray}
n^2&=&(1-a)^2b^2\,,\\
(\nu+n)^2&=&(1+a)^2b^2\,.
\end{eqnarray}
There are two types of solutions to these:
\begin{equation}
a={\nu\over 2n+\nu},\ \ \ \ b=\pm {2n+\nu\over 2},\label{sol1}
\end{equation}
and
\begin{equation}
a={ 2n+\nu\over \nu},\ \ \ \ b=\pm {\nu\over 2}\,.\label{sol2}
\end{equation}
The divergent part of the action can now be calculated for both
solutions. It turns out to be independent of $C(\theta)$
\begin{equation}
S\sim 2a^2b^2\Lambda\,.
\end{equation}
For the two possible solutions
\begin{eqnarray}
S_1&=&{1\over 2}\,\nu^2\,\Lambda\,,\label{act1}\\
S_2&=&{1\over 2}\,(2n+\nu)^2\,\Lambda\,.\label{act2}
\end{eqnarray}
Thus for even values of $\nu$, which correspond to
't\,Hooft-Polyakov monopoles, the action is minimized by choosing
the solution eq.(\ref{sol2}) and $n=-\nu/2$, and it is UV finite.
On the other hand in the case of interest to us, $\nu=1$, the
action is minimized with eq.(\ref{sol1}) for any $n$ as well as
for eq.(\ref{sol2}) with $n=0$ and $n=-1$, but is UV divergent.

Note that for $\nu=1$ a simple representative of the class of
configurations with minimum action is given by
 \begin{equation}
C(\theta)={1\over 2}(1-\cos\theta),\ \ \ A=0,\ \ \ B=0\,.
\end{equation}
This is precisely the abelian Dirac monopole solution\footnote{
Although our minimization procedure has only established that $A$
and $B$ vanish at the origin, it is a straightforward matter to
show that in this case they will also vanish at all values of $r$.
Any nontrivial $r$-dependence of either $A$ or $B$ immediately
increases the energy due to the square of the last two terms in
eq.(\ref{field}).}.

 We conclude from this discussion that a classical configuration that minimizes the
action in the unitary gauge is the abelian Dirac monopole
\begin{equation}
b^a_\mu={1\over g}\  \delta^{a3}(1-\cos\theta)\
\partial_\mu\phi\,.
\end{equation}
The action for this configuration is the same as in QED, as all
the nonabelian components of the gauge field vanish. The VEV of
the vortex operator in the Georgi-Glashow model in the leading
perturbative order is thus
\begin{equation}
\left< V \right>_0=e^{-{\pi\over 2g^2}\,\Lambda}\,.
\end{equation}

Our next step is to consider the one loop corrections. To this end
we write the vector potential as the sum of the classical solution
and the fluctuation field
\begin{equation}
A^a_\mu=A^a_{{\rm cl}\mu}+a^a_\mu\,,
\end{equation}
with $A^a_{{\rm cl}\mu}$ defined in eq.(\ref{classg}). In this
expansion we consider the classical field to be of order $1/g$,
while the fluctuation field to be of order one. The Higgs field is
also considered to be of order one. For a one loop calculation we
need to expand the action in eq.(\ref{expvgg}) to order one. Note
that the relation between the classical solution and the Higgs
field is such that
\begin{equation}
D^{ab}_\mu(A_{\rm cl})\hat\phi^b=0\,. \label{covd}
\end{equation}
Writing the Higgs field in terms of its modulus $\rho$ and the
unit vector $\hat\phi^a$ we find that
\begin{equation}
D^{ab}_\mu(A_{\rm cl})\phi^b=\hat\phi^a\partial_\mu\rho\,.
\end{equation}
Thus to order one the action in eq.(\ref{expvgg}) is
\begin{equation}
{\,\pi\over 2g^2}\Lambda+{1\over 2}(D^{ab}_{[\mu}(A_{\rm
cl})a^b_{\nu]})^2+\epsilon^{abc}F^a_{\mu\nu}(A_{\rm cl})a^b_\mu
a^c_\nu+(\partial_\mu\rho)^2+M^2\rho^2
\end{equation}

Although the classical vector potential $A_{\rm cl}$ depends on
the direction of the field $\phi$, the integral over $a^a_\mu$ in
eq.(\ref{expvgg}) does not depend on it. Thus the integration over
$a^a_\mu$ and $\phi$ factorizes. The part of the action that
contains $\phi$ does not know anything about the monopole field
and thus is unaffected by the presence of the vortex operator in
the path integral.

One should be a little careful with the analysis just presented,
as it involves a perturbative calculation in the glue sector. This
part of the calculation is in fact infrared divergent. The
integration over the vector potential $a^a_\mu$ is very similar to
the integration over the Higgs field in the QED case --- both are
charged fields coupled to a monopole. Thus the result of the
integration over $a^a_\mu$ will be formally similar to
eq.(\ref{vev1}), except that the power in the last factor may be
somewhat different. The important difference though is that the
vector field  is massless, and so instead of $M$ as in
eq.(\ref{vev1}), here we will in fact have zero. This zero is the
infrared divergence of the one loop correction to the monopole
action.

This divergence however does not invalidate our conclusion. An
expectation value of any local operator is not calculable
perturbatively in a Yang-Mills theory, because it inevitably
involves the knowledge of the infrared modes which are
nonperturbative. However one can certainly calculate
perturbatively a correlation function of any two such operators as
long as the separation between them is smaller than the inverse
coupling constant. If instead of an expectation value
$\left<V\right>$ we consider the correlation function
$\left<V^*(x)V(y)\right>$, the infrared cutoff in the calculation
of the one loop correction will be provided by the separation
$|x-y|$. On the other hand this will not change any of the
important features of our calculation. The leading classical
configuration will still be Abelian --- this time an Abelian
monopole-antimonopole pair. For this classical background
eq.(\ref{covd}) still holds, and thus the Higgs field decouples
from the background. The integration over the Higgs field
therefore does not involve the background and the result knows
nothing about $|x-y|$
\begin{equation} \left<V(x)V^*(y)\right>=e^{-{\pi\over
g^2}[\Lambda-{1\over|x-y|}]}\left({1\over\Lambda^2(x-y)^2}\right)^\gamma\,,
\end{equation}
where the constant $\gamma$ is determined by the integration over
the fluctuations of the gluon field. Thus the vortex-antivortex
correlation function is not affected by the Higgs field to one
loop order, as long as perturbation theory can be applied, i.e.
$g^2\ll M$, $g^2\ll|x-y|^{-1}$.

We conclude this discussion with the observation that in the
infinite Higgs mass limit, the Higgs field can be integrated
exactly in the path integral eq.(\ref{expvgg}). In this limit the
Higgs integral is dominated by the field configurations with
vanishingly small action. The only appearance of the Higgs then is
in the source term $\hat\phi^as_{\mu\nu}$. The Higgs integral then
degenerates into the form that allows the explicit calculation
using the expression for the so-called
Harish-Chandra-Itzykson-Zuber integral
\cite{HarishChandra},\cite{Itzykson} over the unitary group
$U(N)$:
\begin{equation}
\int_U[dU] e^{{1\over g}{\rm tr}(AUBU)} =
(\prod_{n=1}^{N-1}n!)(\frac{1}{g})^{{-\frac{N(N-1)}{2}}}
\frac{\det||e^{\frac{1}{g}a_ib_j}||}{\Delta(a)\Delta(b)}\,,
\end{equation}
where $DU$ is the Haar measure,  $A$ and $B$ are hermitian
$N\times N$
 matrices,   $\Delta(a)$ and $\Delta(b)$ are Vandermonde
determinants expressed in terms of the corresponding eigenvalues
$a_i$ and $b_i$
\begin{equation}\label{eq:Vd}
\Delta(a) = \prod_{1<i<j<N}(a_i - a_j)\,.
\end{equation}

Exploiting  the representation (\ref{eq:unh}) for the unit vector
Higgs field  one can integrate  over $\hat\phi^a$ in
eq.(\ref{expvgg})
\begin{equation}\label{eq:expex}
\left<V\right> = Z^{-1}\int [dA_\mu]\exp-\int d^3x\ {1\over
4}\left\{ F^2 + s^2\left( 1 - {g^2\over2\pi^2\Lambda}\ln \ \left[
\frac{4\pi\sinh{\cal{F}}}{{\cal{F}}}\right]\right)\right\}\,,
\end{equation}
where
\begin{equation}\label{eq:FK}
{\cal{F}}= {4\pi\over g\Lambda }\, \sqrt{\tilde
F^a_{\mu}\tau_{\mu}(C)\tilde F^a_{\nu}\tau_{\nu}(C)}\,.
\end{equation}
In eq.(\ref{eq:expex}) the UV cutoff $\Lambda$ is understood as the
inverse of the discretization scale. This representation for  the vacuum
expectation value of vortex operators can be further simplified, by noting
that finite contributions to the path integral come only from
configurations for which $F\sim (\Lambda^2/g)$, since only those have a
chance of cancelling the UV and IR divergence coming from the $s^2$ term.
For these fields only the positive exponent in $\sinh$ has to be kept.
Thus we arrive at
\begin{equation}
\left<V\right> = Z^{-1}\int [dA_\mu]\exp-\int d^3x\ {1\over
4}\left\{ F^2 + s^2\left( 1 - {g^2\over 2\pi^2\Lambda}\left[
{\cal{F}}-\ln {{\cal{F}}\over 2\pi} \right]\right)\right\}\,.
\end{equation}
This is the explicit gauge invariant  expression for the vortex
operator in pure gluodynamics in 2+1 dimensions.

\section{Conclusions}

In this paper we have investigated the influence of dynamical
matter fields on the ultraviolet behaviour of the vacuum
expectation value and correlator of vortex operators. We have done
this for two models in 2+1 dimensions and have seen very different
effects. For noncompact QED with a charged scalar field we have
seen how the matter fields induces power like factors in the VEV
and correlator. Thus the ultraviolet behaviour of the vortex
operator in the theory with dynamical Higgs is different from that
in the theory without the Higgs particle at distance scales
$|r|<M^{-1}$. In contrast, for the $SU(2)$ gauge theory with the
Higgs field in the adjoint representation, we have seen that the
matter fields have no effect at one-loop order in the perturbative
regime. In this theory, therefore, for $M\gg g^2$ the vortex
operator is insensitive to the presence of the dynamical Higgs as
soon as $|r|\ll g^{-2}$.

This is somewhat surprising, since generically one expects the
presence of extra degrees of freedom to affect all observables. We
may speculate that this is possibly related to another unexpected
observation that holds in the same theory. Namely, it has been
noted \cite{mike} that the spectrum of the $2+1$ dimensional
$SU(2)$ gauge theory with adjoint Higgs is separated into two very
distinct parts. One part contains glueballs, whose masses are
almost completely independent of the Higgs mass as long as the
Higgs is heavier than the gauge coupling. The other part of the
spectrum contains ``bound states" of the Higgs boson, and scales
appropriately with the Higgs mass. The fact that the glueballs are
not affected by the mass of the Higgs does not have a simple and
natural explanation. Our finding is akin to this effect and
suggests that the vortex operator correlation functions are
heavily dominated by the glueball intermediate states.

\acknowledgments A.~Kh.~thanks the School of Mathematics and
Statistics at the University of Plymouth for their kind
hospitality.

\appendix

\section{Appendix}

In this appendix we perform the minimization of the UV divergent
part of the monopole action:

\begin{equation}\label{eq:enfun}
    {\cal S}^{UV}=\int {d\theta \over \sin\theta}
\left( (C'-AB)^2 + (B(C+n)+A')^2 \right)\,.
\end{equation}
We now keep the function $C(\theta)$ fixed and minimize this
functional with respect to $A$ and $B$:
\begin{eqnarray}
\frac{\delta {\cal S}^{UV} }{\delta A} &=&
\frac{B(C'-AB)}{\sin\theta}+
\frac{d}{d\theta}\left[\frac{A'+B(C+n)}{\sin\theta}\right]=0\,,\label{eq:veq1}
\\
\frac{\delta {\cal S}^{UV} }{\delta B} &=&
A(C'-AB)-(C+n)(A'+B(C+n))=0\,.\label{eq:veq2}
\end{eqnarray}
Expressing $B$   from eq.(\ref{eq:veq2})
\begin{equation}\label{eq;B}
    B = \frac{AC'- A(C+n)}{A^2+(C+n)^2}\,,
\end{equation}
and substituting this expression into the first equation
(\ref{eq:veq1}) we find after some simple algebra:
\begin{equation}\label{eq:A}
\frac{(C+n)(AC'-A'(C+n))}{A^2+(C+n)^2}+A'=
A\frac{C'(C+n)+AA'}{A^2+(C+n)^2}\,.
\end{equation}
Now defining
\begin{equation}\label{eq:A1}
Y= \frac{d}{d\theta}\ln\left[ A^2+(C+n)\right]\,,
\end{equation}
we can write (\ref{eq:A}) in the form of a differential equation
for the unknown function $Y(\theta)$:
\begin{equation}\label{eq:Y}
    \frac{dY}{d\theta} + \frac{1}{2}\ Y^2 -\cot\theta\ Y = 0\,.
\end{equation}
In order to integrate this non-linear differential equation we
introduce a new  function $Z$ defined by
\begin{equation}\label{eq:Z}
    Z= \cot\theta\ Y\,,
\end{equation}
for which the equation $(\ref{eq:Y})$ reduces to the simple form
\begin{equation}\label{eq:Z1}
 \frac{dZ}{d\theta} +( Z+\frac{1}{2}\ Z^2)  \tan\theta  = 0\,.
\end{equation}
The general solution to this equation is
\begin{equation}\label{eq:solZ}
Z(\theta)= \frac{2a\cos\theta}{1-\cos\theta}\,,
\end{equation}
where $a$ is an integration constant. From this solution  one
finds $\ln\left[ A^2+(C+n)\right]$  by integrating $Y$ in
$(\ref{eq:A1})$ :
\begin{equation}\label{eq:Aa}
    A^2+(C+n)^2=(1-a\cos\theta)^2 b^2\,,
\end{equation}
where $b$ is a new constant of integration. Note that the
constants $a$ and $b$ are subject of the equations that follows
from the boundary conditions eqs.(\ref{bound1}) and
(\ref{bound2}).


\end{document}